\documentclass[reprint,amsmath,amssymb,aps]{revtex4-2}

\usepackage{graphicx}
\usepackage{dcolumn}
\usepackage{bm}
\usepackage{amssymb}

\begin{document}

\title{General Spatial Photonic Ising Machine Based on Interaction Matrix Eigendecomposition Method}

\author{Shaomeng Wang, Wenjia Zhang*, Xin Ye and Zuyuan He}

\affiliation{State Key Laboratory of Advanced Optical Communication Systems and Networks \\
Shanghai Jiao Tong University, Shanghai 200240, China\\
Email: wenjia.zhang@sjtu.edu.cn}

\date{\today}

\begin{abstract}
The spatial photonic Ising machine has achieved remarkable advancements in solving combinatorial optimization problems.
However, it still remains a huge challenge to flexibly mapping an arbitrary problem to Ising model.
In this paper, we propose a general spatial photonic Ising machine based on interaction matrix eigendecomposition method.
Arbitrary interaction matrix can be configured in the two-dimensional Fourier transformation based spatial photonic Ising model by using values generated by matrix eigendecomposition.
The error in the structural representation of the Hamiltonian decreases substantially with the growing number of eigenvalues utilized to form the Ising machine. 
In combination with the optimization algorithm, as low as $\sim$65\% of the eigenvalues is required by intensity modulation to guarantee the best probability of optimal solution for a 20-vertex graph Max-cut problem, and this probability decreases to below $\sim$20\% for zero best chance.
Our work provides a viable approach for spatial photonic Ising machines to solve arbitrary combinatorial optimization problems with the help of multi-dimensional optical property.
\end{abstract}


\maketitle

\section{Introduction}

The mathematical model describing the interaction and phase transition behavior of large-scale complex systems, called the Ising model \cite{RevModPhys.39.883,McCoyWu+1973}, has found its important applications in solving combinatorial optimization problems derived from economics~\cite{HOSSEINY2019644}, climatology \cite{Ma_2019}
and epidemiology~\cite{MELLO2021125963}.
The minimum Hamiltonian, corresponding to the solution for these applications,  can be expressed as $H(\boldsymbol{x})=-\boldsymbol{x}^T\boldsymbol{J}\boldsymbol{x}-\boldsymbol{h}^T\boldsymbol{x}$, where $\boldsymbol{J}$ is the interaction matrix, $\boldsymbol{h}$ is the external field vector and $\boldsymbol{x}$ is the spin state  \cite{mohseni2022ising}.
Currently, a huge challenge for conventional electrical hardware to run Ising model is that the required computation power grows exponentially with the problem scale, which leads to research opportunities being framed as next-generation computing technology such as photonic Ising machine \cite{marandi2014network,cen2022large,Lu:23,pierangeli2019large,PhysRevLett.127.043902,Sun:22,Ye:23}. 
There are several proposals for constructing photonic Ising machine based on different physical principles and devices, mainly including coherent photonic Ising machine (CIM) \cite{marandi2014network,cen2022large,Lu:23} and spatial photonic Ising machine (SIM) \cite{pierangeli2019large,PhysRevLett.127.043902,Sun:22,Ye:23}. 

Compared with coherent photonic Ising machine that a network of artificial spins are running as a pulse train in the degenerate optical parametric oscillator, spatial photonic Ising machine uses spatial light modulators (SLMs) to encode spins in the phase and amplitude, then performs a two-dimensional optical Fourier transformation to realize all-at-once spin interactions by multi-beam laser interference \cite{pierangeli2019large}.
Therefore, spatial photonic Ising machine, instead of realizing minimum energy oscillation system that maps global or local optimal solution, is to construct a large-scale physical Ising model using optical Fourier transformation with optimization process participated by electrical hardware.
The Hamiltonian of spatial photonic Ising machine can be expressed as \cite{pierangeli2019large}
\begin{equation}
    H = -C \cdot \boldsymbol{x}^T\boldsymbol{\xi}\boldsymbol{\xi}^T\boldsymbol{x}
    \label{ising1}
\end{equation}
where $C$ is a constant, $\boldsymbol{x}$ denotes spin vector given by optical phase and $\boldsymbol{\xi}$ denotes vector given by optical amplitude to compose interaction matrix given by
\begin{equation}
   \boldsymbol{J}\propto\boldsymbol{\xi}\boldsymbol{\xi}^T.
    \label{interaction}
\end{equation}

Eqs.~(\ref{ising1}) and (\ref{interaction}) show that SIM has excellent scalability, and more importantly, full connectivity, allowing connections between any spin and any other spin defined by spatial amplitude.
For the one, scalability is provided by light manipulation in the two-dimensional space. For instance, a 1080p resolution SLM can modulate 1920$\times$1080 spins to form a million-scale Ising model.
For another, full connectivity, not a necessity in real physics, is a key property that facilitates arbitrary optimization problem mapping.
But full connectivity is very hard to be emulated physically by using oscillation systems as extra delay network needs to be built to interfere two pulses with interaction amplitude of $\boldsymbol{J}$.

However, Eq.~(\ref{interaction}), on the other hand, also implies major limitation of SIM for mapping interaction matrix, which can be explained by mathematical difficulty of constructing a two-dimensional matrix through multiplying a one-dimensional vector and its transposition. 
This limitation confines computing capability of SIM that is inherently only able solve one-dimensional problems such as subset-sum problem \cite{huang2021antiferromagnetic}. 
And interaction matrix $\boldsymbol{J}$ from most combinatorial optimization problems is difficult to satisfy the form of Eq.~(\ref{interaction}) \cite{10.3389/fphy.2014.00005}, which leads to the inability of the spatial photonic Ising machine in facing the majority of combinatorial optimization problems.
Moreover, according to Eq.~(\ref{ising1}), the external field vector $\boldsymbol{h}$ in the Ising model cannot be realized neither, but if an Ising machine with arbitrary interaction matrix can be implemented, any $\boldsymbol{h}$ can be brought into it at the cost of adding one spin.
Therefore, how to create the general spatial photonic Ising machine become a very critical issue for this technology towards real applications.
Recently, methods such as optical matrix multiplication \cite{ouyang2023ondemand}, wavelength-division multiplexing \cite{luo2023wavelengthdivision} and low-rank matrix construction \cite{yamashita2023spatialphotonic} have been proposed to extend the generality of SIM, but modulation and quantitative analysis of the partial composition of interaction matrices, which is highly relevant to the mapping and solving of practical problems, have not been mentioned.
Besides that, the above work starts with directly optimizing the structure of the spatial photonic Ising machine, thus increasing the configurability, rather than realizing arbitrary interactions of Ising machines through top-level physical implementations.
In our previous work, we have improved the flexibility of the spin configuration by introducing spatial resolved quadrature modulation in the SLM, and have further realized arbitrary intensity modulation based on Euler equation in a single SLM \cite{ Sun:22, Ye:23}.
These works focus on the flexible spin interaction configuration in a spatial photonic Ising machine to reduce the difficulty of experimental implementation. 

In this paper, we propose a general spatial photonic Ising machine (G-SIM) based on interaction matrix eigendecomposition method, where intensity vectors can be modulated in parallel enabled by multi-dimensional optical property.
We concentrate on the two-dimensional interaction matrix, the core of Ising's problem, and introduce eigendecomposition, a polynomial-complexity procedure to obtain intensity vectors, in order to accomplish the configuration of an arbitrary interaction matrix using the existing SIM structure.
The impact of modulating different numbers of intensity vectors, including the matching degree of Hamiltonian of the Ising model and solution divergence for the same optimization problem, is quantitatively analyzed, showing that linear fit can be achieved for Ising models of different sizes and connectivity.
Furthermore, for solving the Max-cut problems with a random weighted 20-vertex graph, as low as $\sim$65\% of the eigenvalues is required by intensity modulation to guarantee the best probability and this probability decreases to below $\sim$20\% for zero best chance.
\begin{figure*}[tb]
    \centering
    \includegraphics[width=0.65\linewidth]{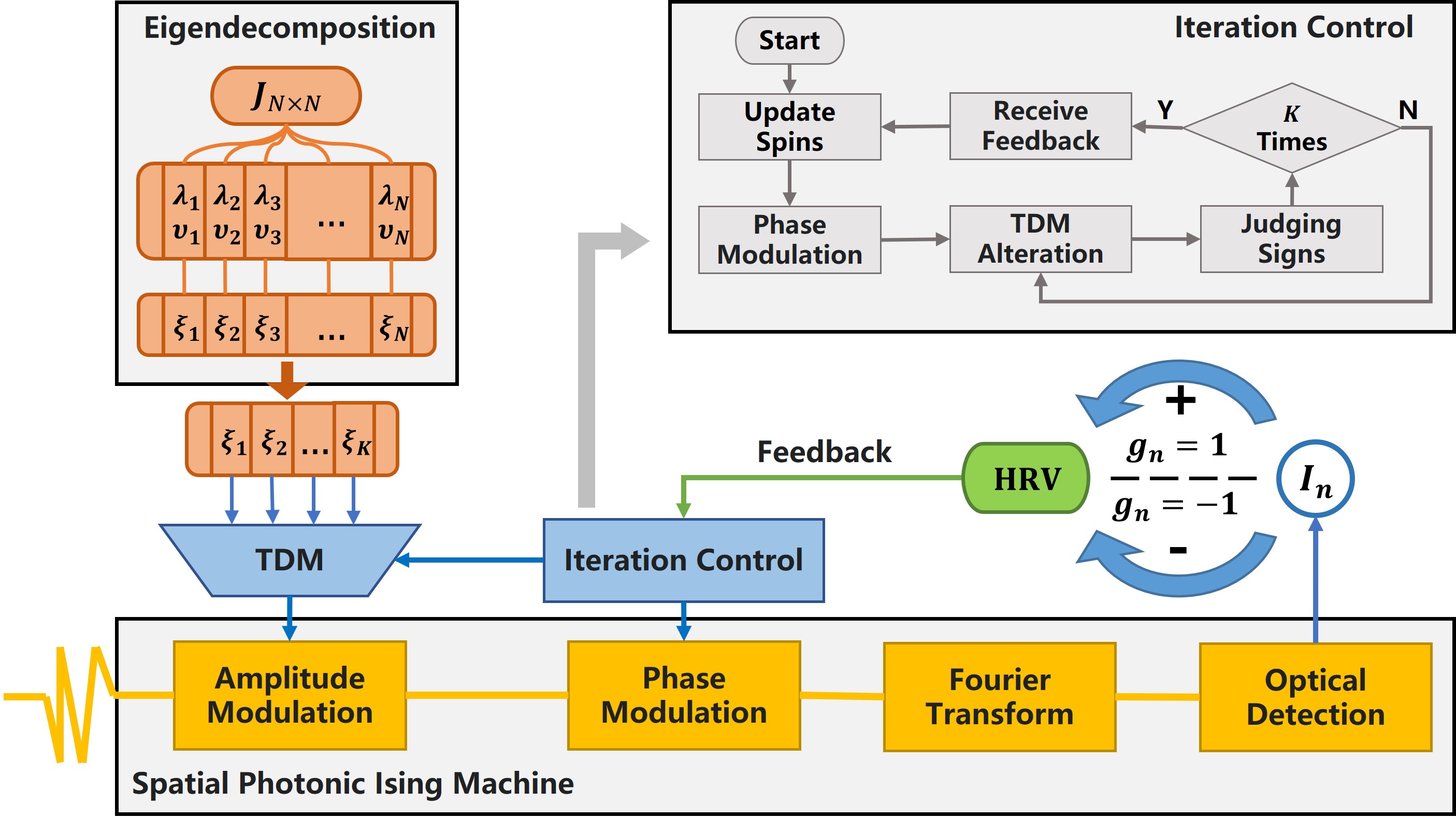}
    \caption{\label{blockdia} Architecture and workflow of the time-division multiplexing (TDM)-based general photonic Ising machine, where the part beyond the spatial photonic Ising machine (SIM) is performed using electrical hardware.
    Some external factors affecting amplitude modulation and phase modulation, such as the line and alignment of the beam, are neglected here for a better illustration of the principle. Meanwhile, the spatial-Euler Ising machine from our previous work \cite{Ye:23} is utilized to achieve arbitrary amplitude modulation, which is based on only one phase-only SLM. Phase modulation is also directly realized by one phase-only SLM.
    }
\end{figure*}

\section{Principle of general spatial photonic Ising machine}
For practical applications of solving combinatorial optimization problems, the interaction matrix from problem mapping algorithms is typically arbitrary, with unknown connectivity pattern for an Ising model.
Therefore, the interaction matrix often fails to satisfy the condition  $\boldsymbol{J}\propto\boldsymbol{\xi}\boldsymbol{\xi}^T.$
The lack of satisfaction in the form of the interaction matrix can be addressed by considering matrix decomposition techniques. 

Fig.~\ref{blockdia} illustrates the architecture and workflow of the general spatial photonic Ising machine based on interaction matrix eigendecomposition method, which can be enabled by time-division multiplexing (TDM) modulation scheme.
As shown in Fig.~\ref{blockdia}, a typical spatial photonic Ising machine is composed of amplitude modulation, phase modulation, Fourier transform and optical detection \cite{pierangeli2019large}.
The intensity value at the center point after optical detection can be expressed as 
\begin{equation}
    I = \sum_{ij} \xi_i \xi_j e^{\varphi_i} e^{\varphi_j}
    \label{inten}
\end{equation}
where $\xi_i$ ($\xi_j$) and $\varphi_i$ ($\varphi_j$) are the amplitude and phase of the optical signal at pixel position \emph{i} (\emph{j}) on the SLM, respectively. The phase $\varphi$ to $0$ or $\pi$, corresponding to the respective spin state $x$, is updated timely depending on optimization algorithms. 
According to the mathematical expression of Eqs. (\ref{ising1}) and (\ref{inten}), the detected light is numerically proportional to the Hamiltonian of the Ising model $H \propto - I$.

Interestingly, eigenvalue decomposition, a common matrix decomposition method, can decompose any diagonalizable matrix $\boldsymbol{J}$, given by an arbitrary interaction matrix, into eigenvectors $\boldsymbol{\upsilon}$  and corresponding eigenvalues $\lambda$ \cite{golub2013matrix}.
Conversely, the matrix $\boldsymbol{J}$ can be represented using eigenvalues and eigenvectors, as shown below
\begin{widetext}
\begin{equation}
    \boldsymbol{J}=\boldsymbol{Q \Lambda Q^T}=
    \begin{pmatrix}
        \upsilon_{11} & \upsilon_{21} & \cdots &\upsilon_{N1} \\
        \upsilon_{12} & \upsilon_{22} & \cdots &\upsilon_{N2} \\
        \vdots & \vdots & \ddots & \vdots \\
        \upsilon_{1N} & \upsilon_{2N} & \cdots &\upsilon_{NN}
    \end{pmatrix}
    \begin{pmatrix}
        \lambda_1 & 0 & \cdots & 0 \\
        0 & \lambda_2 & \cdots & 0 \\
        \vdots & \vdots & \ddots & \vdots \\
        0 & 0 & \cdots & \lambda_N
    \end{pmatrix}
    \begin{pmatrix}
        \upsilon_{11} & \upsilon_{21} & \cdots &\upsilon_{N1} \\
        \upsilon_{12} & \upsilon_{22} & \cdots &\upsilon_{N2} \\
        \vdots & \vdots & \ddots & \vdots \\
        \upsilon_{1N} & \upsilon_{2N} & \cdots &\upsilon_{NN}
    \end{pmatrix}^T
    \label{decom}
\end{equation}
\end{widetext}
where $\boldsymbol{\Lambda}$ is the diagonal matrix of eigenvalues $\lambda_n$, and the respective corresponding eigenvectors $\boldsymbol{\upsilon}_n$ are the columns of $\boldsymbol{Q}$.
As one of the commonly employed matrix decomposition methods, eigendecomposition has the polynomial level of time complexity. Consequently, in comparison to the exponential level combinatorial optimization process, its time complexity is sub-magnitude and fully acceptable.

Taking the interaction matrix $\boldsymbol{J}$ as the aforementioned decomposed square matrix, it requires multiple modulations and aggregated processing of the optical intensity information for a spatial photonic Ising machine according to Eqs.~(\ref{inten}) and (\ref{ising1}).
To achieve this, one possible solution, called time-division multiplexing, is to utilize multiple cycles of intensity modulation within a single spatial photonic Ising machine.
By maintaining the same spin state, the obtained intensity values from each modulation are accumulated (or subtracted) according to Eq.~(\ref{decom}). This process allows us to obtain the Hamiltonian Representative Value (HRV) of the target Ising model for that particular state.

In the electronic hardware section, the main tasks include performing eigenvalue decomposition on the interaction matrix and controlling the iterative spin states of the Ising model. 
The computational complexity of eigenvalue decomposition is polynomial, making it feasible to implement for the overall system.
Before implementing the eigendecomposition, we represent the interaction matrix as an $N \times N$ symmetric square matrix $\boldsymbol{J}$, where $N$ is the number of spins of the Ising model as the configuration target, and the values of the $k$-th row, $h$-th column and $h$-th row, $k$-th column of the matrix  $\boldsymbol{J}$ are assigned as half of the interaction intensity between spin $h$ and spin $k$, such that $J_{hk} = J_{kh}$.
With this premise, it is guaranteed that $N$ sets of eigenvalues and eigenvectors consisting of real numbers are obtained after the eigendecomposition of $\boldsymbol{J}$, which is necessary for the subsequent processing.
Taking $\boldsymbol{J}$ as the objective matrix, the decomposition yields the eigenvalues $\lambda$ and eigenvectors $\upsilon$, which satisfy
\begin{equation}
    \boldsymbol{J} = \sum_{n=1}^N \lambda_n \boldsymbol{\upsilon}_n \boldsymbol{\upsilon}_n^T
    \label{equ2}
\end{equation}
deduced from Eq.~(\ref{decom}), and we assume that all the eigenvalues satisfy $|\lambda_1|>|\lambda_2|>...>|\lambda_N|$.
Subsequently, we obtain $N$ sets of intensity vectors given by
\begin{equation}
    \boldsymbol{\xi}_n = P\cdot\sqrt{\left|\lambda_n\right|}\boldsymbol{\upsilon}_n
    \label{equ3}
\end{equation}
where $P$ is a constant related to the pixel division of the SLM.
The positivity or negativity of each eigenvalue needs to be recorded as $g$.
If $\lambda_n$ is positive $g_n$ is recorded as 1, and vice versa is -1.
Suppose that $N$ SIMs are constructed by amplitude modulation based on the above all intensity vector $\boldsymbol{\xi}$, and the individual output intensity 
is obtained according to Eq.~(\ref{inten}).
Furthermore, we determine whether to add or subtract the value based on $g_n$ to obtain the superimposed optical intensity.
After processing the $N$ outputs, the final superimposed intensity is expressed by
\begin{equation}
    I_{SN} = \sum_{n=1}^N g_n I_n = \boldsymbol{x}^T (\sum_{n=1}^N g_n \boldsymbol{\xi}_n\boldsymbol{\xi}_n^T) \boldsymbol{x},
    \label{isn}
\end{equation}
By combining Eqs.~(\ref{equ2}) and (\ref{equ3}), we can get
\begin{equation}
    I_{SN} =  \boldsymbol{x}^T (P^2 \sum_{n=1}^N \lambda_n \boldsymbol{\upsilon}_n \boldsymbol{\upsilon}_n^T) \boldsymbol{x} = P^2 \cdot \boldsymbol{x}^T \boldsymbol{J} \boldsymbol{x}.
    \label{isn2}
\end{equation}
Therefore, given by $H=-\boldsymbol{x}^T \boldsymbol{J} \boldsymbol{x}$, it can be seen from Eq.~(\ref{isn2}) that, for arbitrary interaction matrix $\boldsymbol{J}$, $H \propto -I_{SN}$.
Note that for simplicity, we view the external field vector $\boldsymbol{h}$ as the interaction between an additional spin with the fixed positive state and every other spin, so that $\boldsymbol{h}$ can be treated as one of the columns of $\boldsymbol{J}$, and so the realization of any $\boldsymbol{J}$ in our Ising machine implies any $\boldsymbol{h}$.

\begin{figure}[tb]
    \centering
    \includegraphics[width=0.98\linewidth]{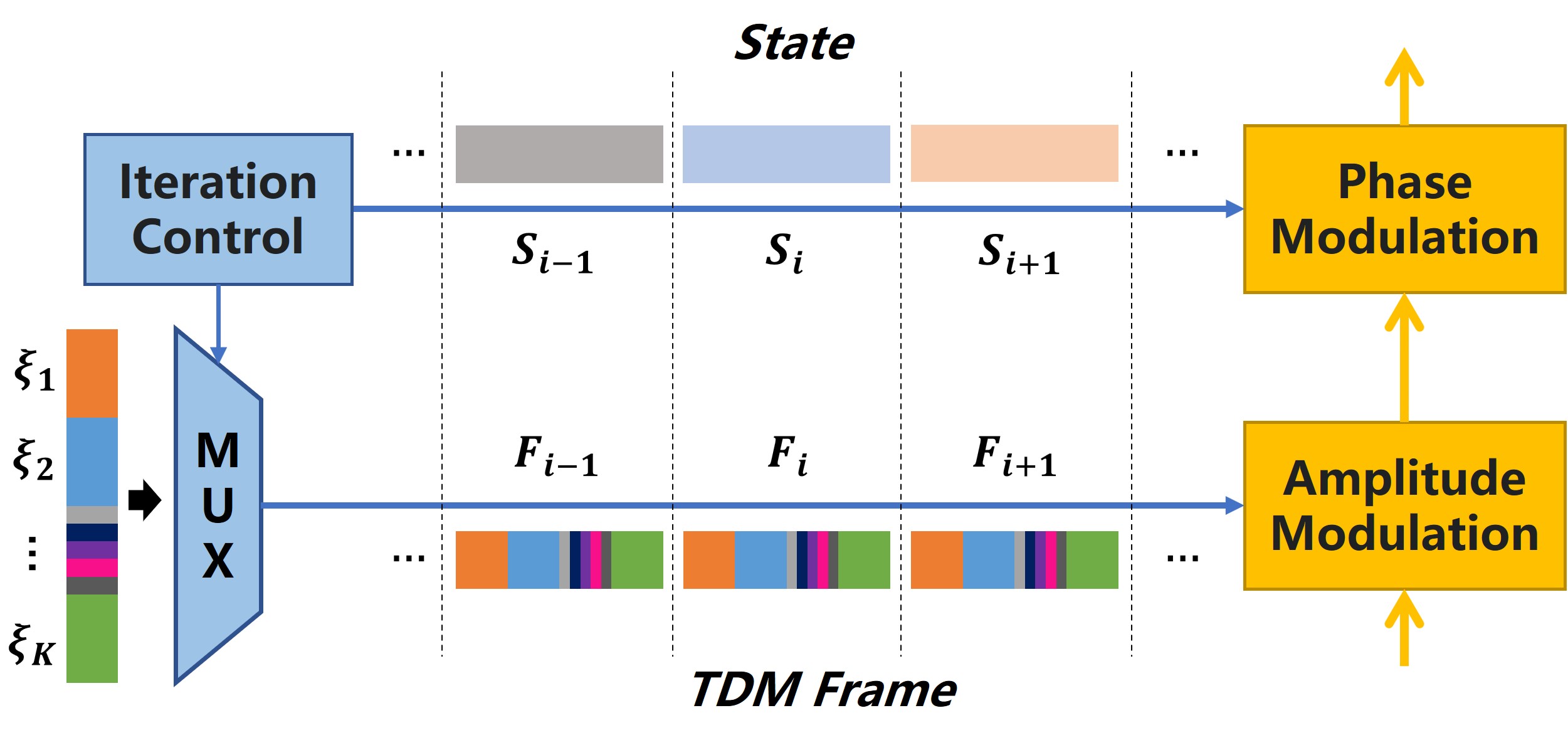}
    \caption{\label{tdm} The schematic of time-division multiplexing (TDM) approach.}
\end{figure}
Through above procedures, a G-SIM can be theoretically constructed, however, with  $N$ sets of intensity vectors required to be amplitude modulated.
As illustrated in Fig.~\ref{tdm}, TDM approach can be used for amplitude modulation in a single SIM instead of recklessly constructing $N$ SIMs.
The iterative control part will carry out the update of the spin states of the Ising machine, and correspondingly, the multiplexer outputs one TDM frame consisting of the intensity vector in each time frame.
Since the mathematical eigendecomposition yields eigenvalues with widely varying absolute values \cite{arnold1967asymptotic}, it is not cost-effective to modulate all intensity vectors.
Therefore we select $K$ of the $N$ intensity vectors for amplitude modulation  instead of all of them, thus obtaining the final superimposed intensity expressed as
\begin{equation}
    I_{SK} = \sum_{n=1}^K g_n I_n, K \leqslant N
    \label{isk}
\end{equation}
which is defined as HRV of the Ising machine.
This approach will reduce the timing and resources pressure on the structure while guaranteeing the computing performance and being consistent with current experimental infrastructure.
Note that mathematically, the significance of the intensity vectors is determined by the absolute magnitude of the corresponding eigenvalues, so we choose $K$ intensity vectors with the largest absolute value of corresponding eigenvalues for the magnitude configuration.


As shown in Fig.~\ref{tdm}, for each spin state, $K$ intensity vectors are required to be modulated in magnitude one by one.
And in order to perform the state optimization, the spin state of the Ising machine is updated by the iterative control part.
It receives feedback on the HRV obtained in the current state and then determines the next state with the chosen optimization algorithm.
Note that we do not impose any restrictions on the specific optimization algorithm used in the structure.
The general flow of the control iteration part is also illustrated in Fig.~\ref{blockdia}.
After G-SIM starts to be applied to the optimization of problems, the algorithm is run to decide the updated spin state taking the value $\boldsymbol{x}$, and in turn the phase modulation is completed.
Further, the multiplexer of TDM is controlled to rotate and output the next intensity vector $\boldsymbol{\xi}_n$ for the amplitude modulation.
The output intensity of SIM is recorded at this point before that value (plus or subtract) is judged by $g_n$.
The HRV of the current state is obtained until $K$ times of amplitude modulation.

\section{Results and Discussion}

\subsection{HRV matching to direct Hamiltonian}

\begin{figure*}[tb]
\centering
    \begin{minipage}[tb]{.42\linewidth}
        \centering
        \includegraphics[scale=0.2]{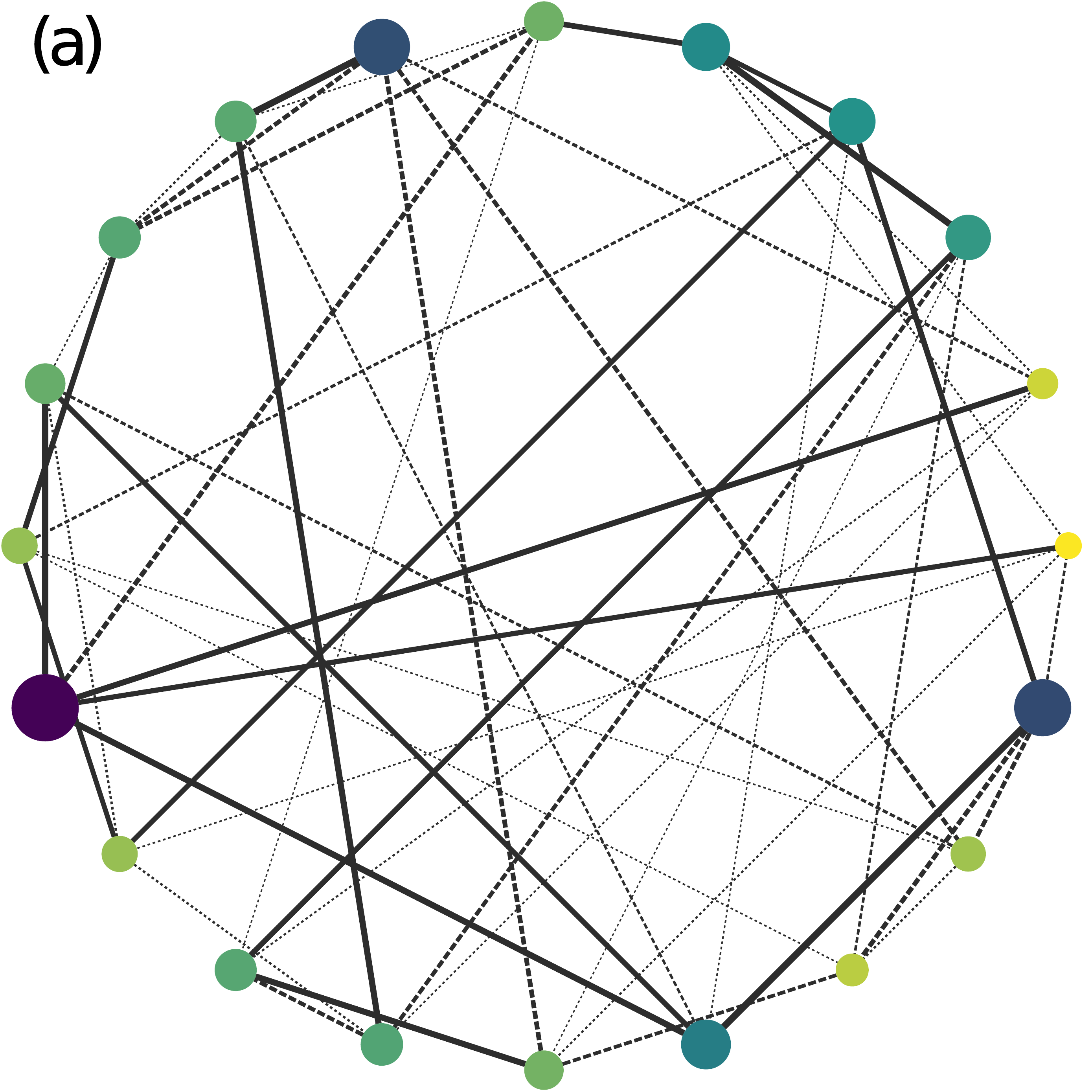}
    \end{minipage} 
    \begin{minipage}[tb]{.57\linewidth}
        \centering
        \includegraphics[scale=0.65]{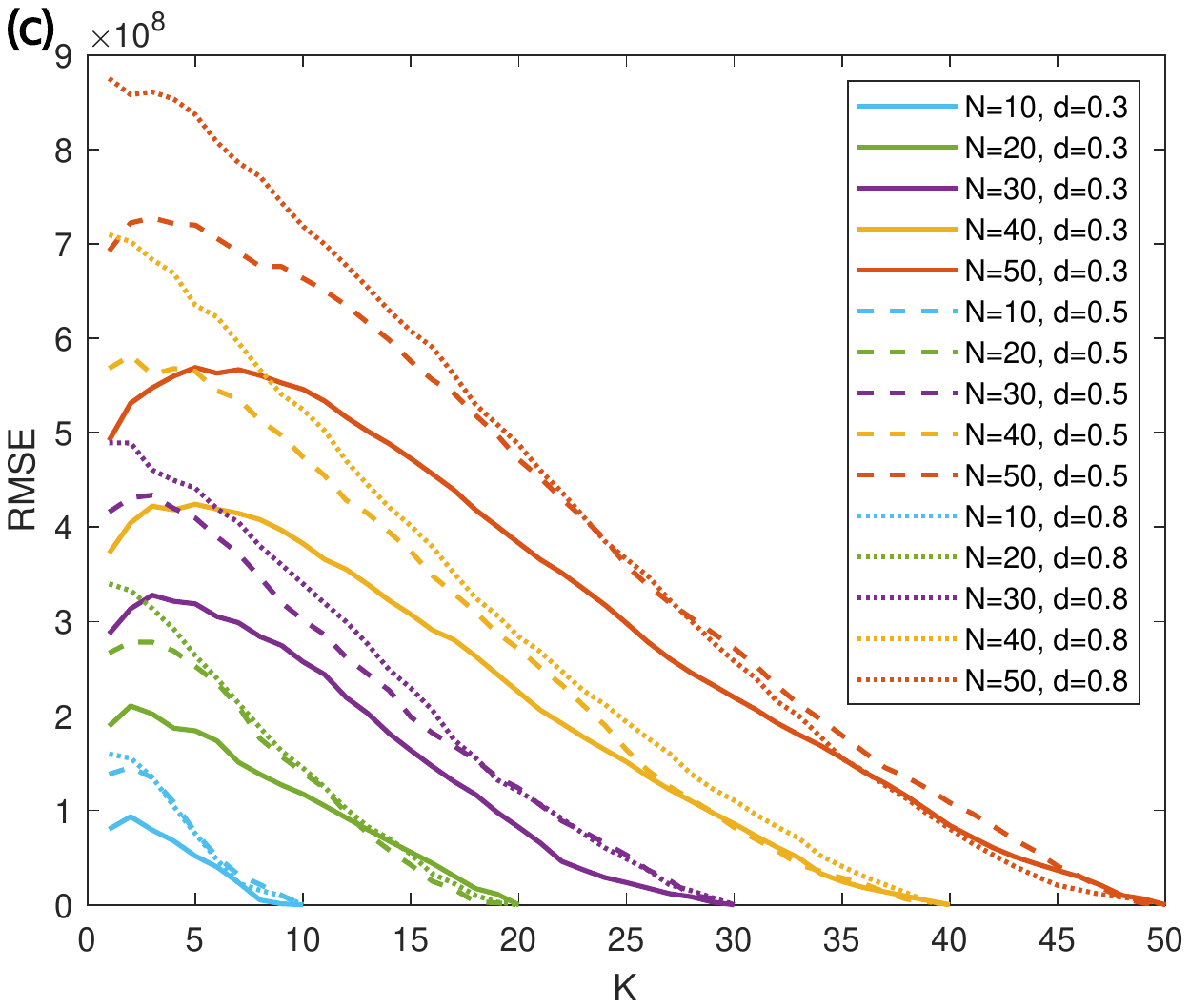}
    \end{minipage} \\
    \quad
    \begin{minipage}[tb]{1\linewidth}
        \centering
        \includegraphics[scale=0.82]{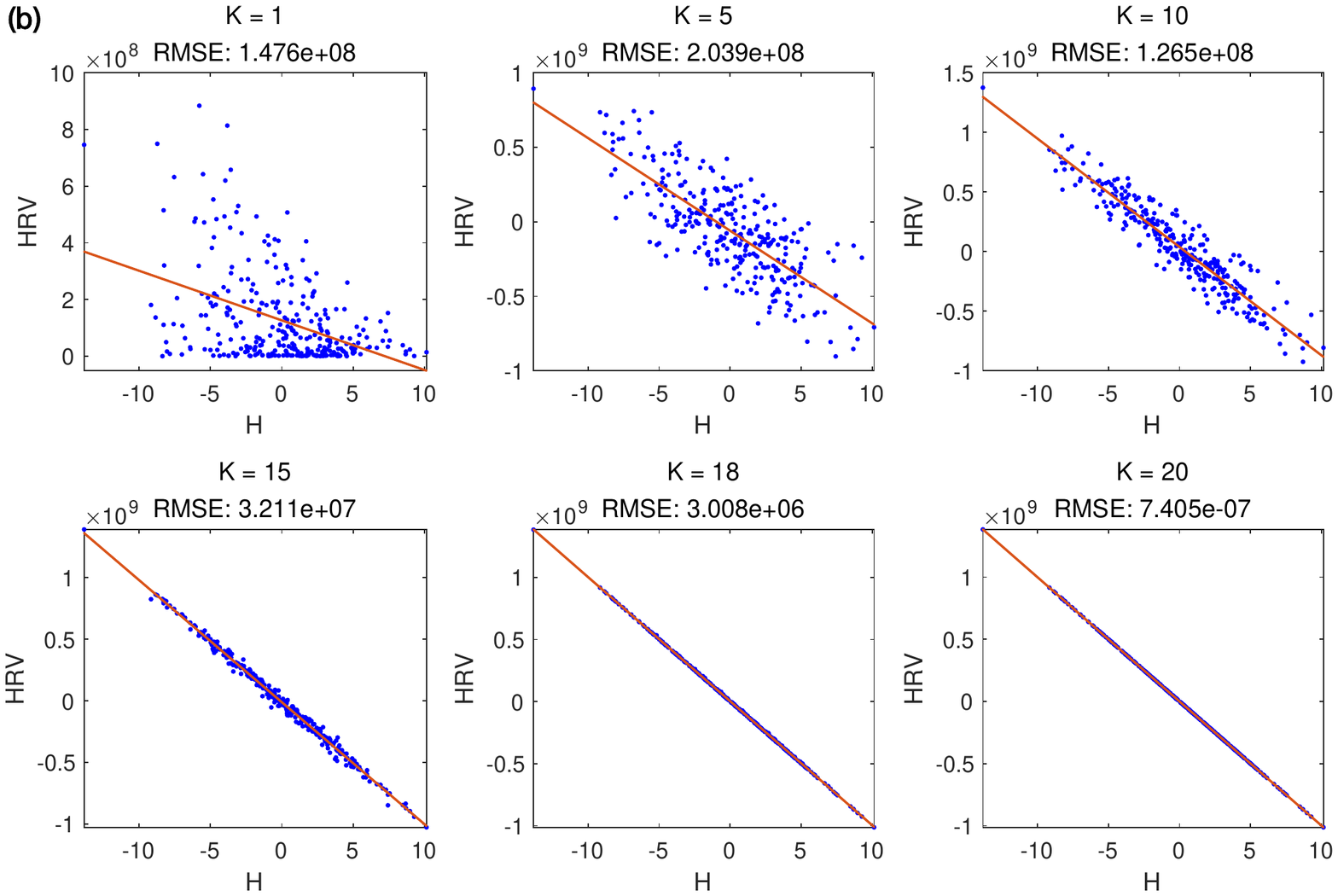}
    \end{minipage} 
    \caption{\label{firgraph} (a) The graph with 20 vertices and degree 5, where the weights are uniformly distributed between 0 and 1. (b) The matching point distributions of HRV and the Hamiltonian for different $K$ values and the results of the linear fit. (c) The trends of RMSE with $K$ for different scales $N$ and graph densities $d$.}
\end{figure*}


For Ising model with the arbitrary interaction matrix and no external field, an equivalent explicit representation is the weighted undirected graph.
Fig.~\ref{firgraph}(a) shows a $20$-vertex graph with degree of $5$, where the edge weights are uniformly distributed between 0 and 1.
This implies that the corresponding Ising model containing 20 spins has the interaction matrix $\boldsymbol{J}$ composed of values with uniformly distribution, i.e., $J_{ij} \sim U(0,1)$.
Furthermore, the Ising model represented by Fig.~\ref{firgraph}(a) can also be configured with different \emph{K} values. 
In this section, 
we focus on the relationship between the obtained HRV and the Hamiltonian calculated directly by $H=-\boldsymbol{x}^T \boldsymbol{J} \boldsymbol{x}$ for different spin states.

Fig.~\ref{firgraph}(b) illustrates the matching results between HRV and the Hamiltonian $H$ obtained by randomly iterating the spin states of the Ising model with different $K$ values.
As $K$ increases, the root mean square error (RMSE), defined as matching degree,  decreases until it approaches zero when $K$ coincides with the number of spins of the Ising model.
Such trend can be observed from the Fig.~\ref{firgraph}(b), where matching degree of data point keeps better as the value of $K$ increases.
At $K = 5$, the approximate linear relationship between HRV and the Hamiltonian has been able to be observed, despite the large error.
When $K=15$, all the matching points have more noticeable linear relationship and the error ratio has been reduced to less than 5\% of the HRV range, and when $K=20$, they have completely linear relationship with zero error.


\begin{figure}[tb]
\centering
    \begin{minipage}[tb]{1\linewidth}
        \centering
        \includegraphics[scale=0.65]{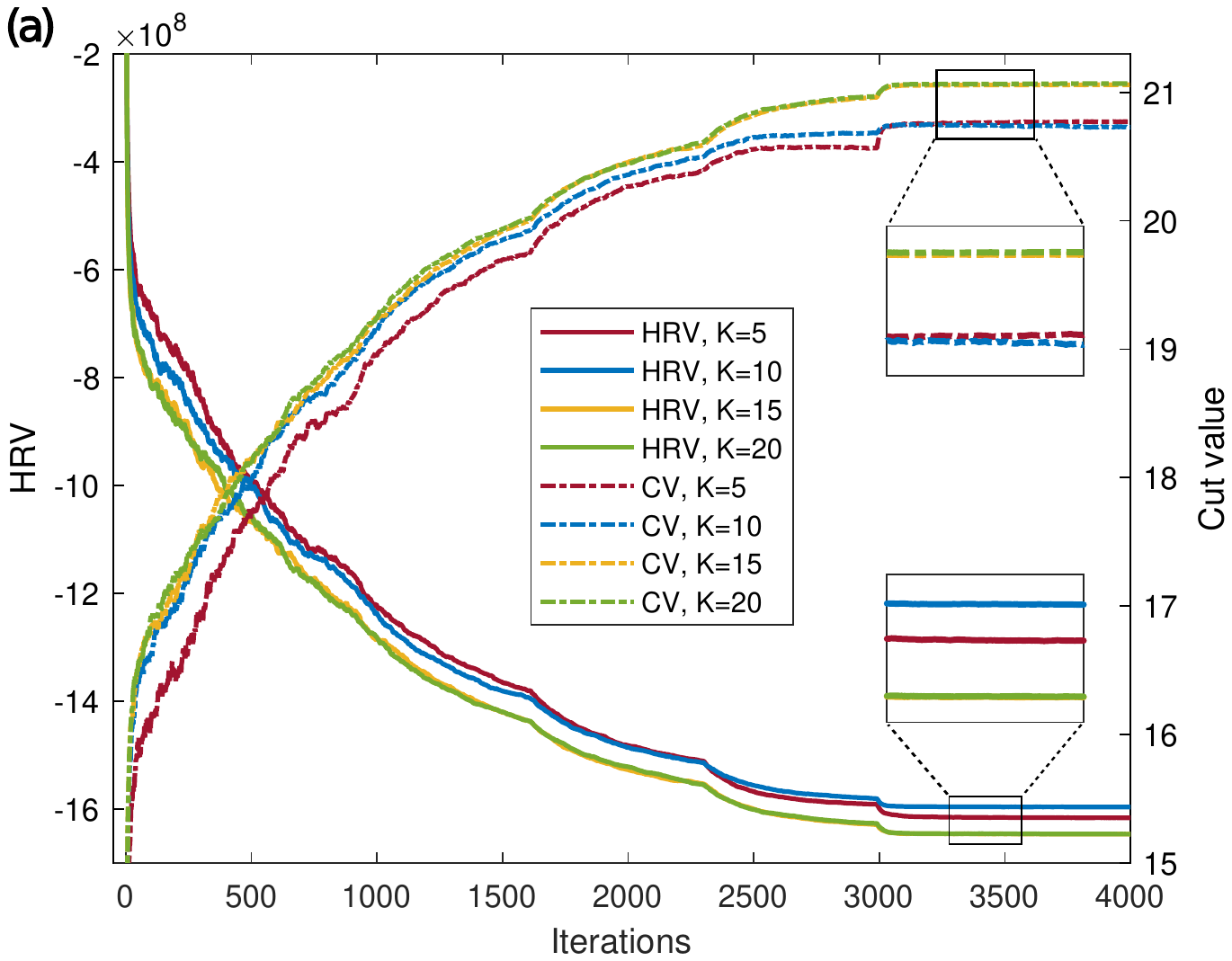}
    \end{minipage} \\
    \quad
    \begin{minipage}[tb]{1\linewidth}
        \centering
        \includegraphics[scale=0.69]{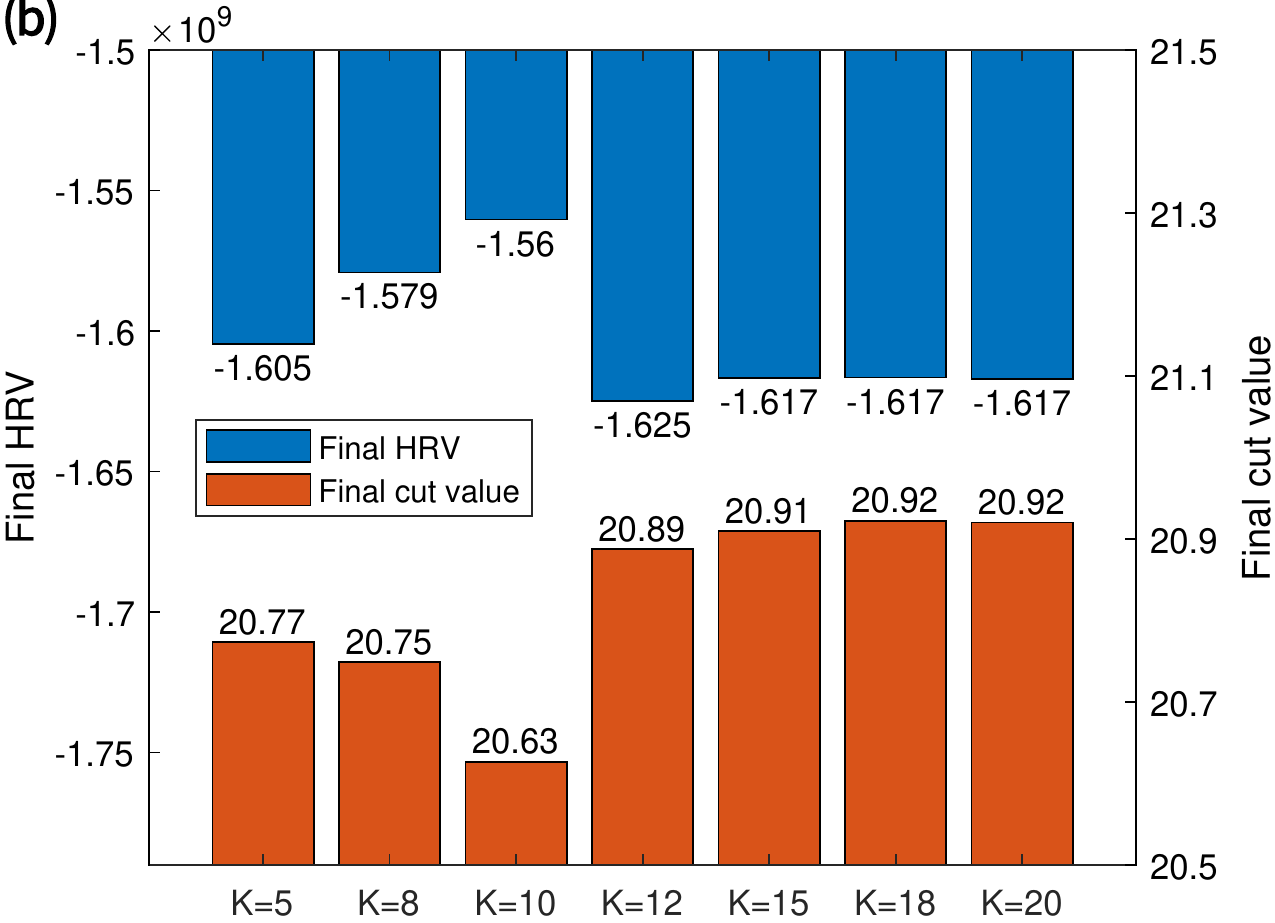}
    \end{minipage} 
    \caption{\label{secgraph} The process of solving the Max-cut problem and the final average result. (a) The trend of HRV and cut values (CV) during simulated annealing for different $K$. (b) The final average HRV and cut values after the optimization was completed for different $K$.}
\end{figure}

The influence of the scale and structure of graphs corresponding to exact Ising model is also considered in the results.
The number of vertices (number of spins) $N$  implies the size of the graph and the Ising model, while for the structure we take the overall graph density
\begin{equation}
    d = \frac{2E}{N(N-1)}
    \label{graphd}
\end{equation}
as the primary impact factor, where $E$ is the number of edges of the graph.
Fig.~\ref{firgraph}(c) shows the variation of the average RMSE with different values of $K$ by configuring the Ising model with different values of $N$ and $d$.
For the same scale, the RMSE exponentially converges to 0 for different graph densities as the $K$ approaches the $N$, but there are some differences in their trends when the $K$ is relatively small.
And for different scales of Ising model, Fig.~\ref{firgraph}(c) presents consistency in the exponential decreasing trend of their RMSE at the same graph density, that is,
\begin{equation}
    RMSE = A \cdot e^{-B (\frac{K}{N} - D)}
    \label{exp}
\end{equation}
where $A$, $B$ and $D$ are constants that vary with $N$ and $d$.
Thus, we believe that the matching relationship is not affected by the size of the graph, but mostly depends on the relative size of $K$ values for $N$ values.

\subsection{Solving Max-cut problem}

For practical problems that can be represented in the form of the Ising model, the Hamiltonian of the Ising model is used as the optimization target for the initial problem.
However, there is uncertainty in the proposed structure with different values of $K$. 
According to the absolute magnitude of the eigenvalues, the error ratio can be estimated as
\begin{equation}
    \mu = 1 - {\frac{\sum_{i=1}^K \lambda_i}{\sum_{i=1}^N \lambda_i} }
    \label{equ5}
\end{equation}
We need to investigate implications of this error on the problem solving performance for many combinatorial  optimization problems, $\textit{e.g.}$, Max-cut problem that defined by following.



Partitioning a given graph into two subsets, such that the sum of weights of the edges connected between them is maximized, is known as the Max-cut problem \cite{GROTSCHEL198123}.
The sum of weights is also known as the cut value, can be expressed as
\begin{equation}
    W = \sum_{<l,k>} w_{l,k} \frac{1-x_lx_k}{2}
    \label{cv}
\end{equation}
where $w_{l,k}$ is the weight of the edge between the $l$-th vertex and the $k$-th vertex, $x_l$ ($x_k$) denotes the subset into which the $l$($k$)-th vertex is divided, and the values of $x$ in the two subsets are 1 and -1, respectively \cite{10.3389/fphy.2014.00005}.
Then, we make $J_{ij} = w_{ij}$ in the corresponding Ising model, and get $W=\sum_{<l,k>}\frac{w_{l,k}}{2}-\frac{H}{2}$, where the value of $x$ that gives $H$ indicates the spin state, so the Max-cut problem can be equivalently transformed into the problem of finding the optimal Hamiltonian for an Ising model with no external fields.
Because of the demand of many practical issues, the weights of undirected graphs applied in Max-cut problems are usually unrestricted and arbitrary.
Furthermore, it is also required to choose an optimization algorithm to solve the problem. In this work, we choose the widely used simulated annealing algorithm \cite{van1987simulated}.

Fig.~\ref{secgraph}(a) shows the average values of HRV and cut values during the iterations for four different $K$ values, with the same optimization algorithm parameters and fixed weights of the graph.
After approximately 3000 state iterations, the HRV and cut values converge to a relatively stable level.
It can be seen from the figure that the average values of final HRV and cut values for $K$ of 20 and 15 are quite close and much better than other cases.
The specific average values of the final HRV and cut values at different $K$ values are further explored.
Fig.~\ref{secgraph}(b) illustrates the obtained average HRV and average cut values after G-SIMs with different $K$ values.
When $K$ is greater than or equal to 15, the performance difference from the final values, both in HRV and cut values, are negligible.
When $K$ is less than 12, the final values start to gradually change. 
However, it is interesting to note that the final values of HRV and cut values for $K=10$ are the worst, and they are instead getting slightly better when $K$ continues to decrease to 8 and 5.

\begin{figure}[tb]
    \centering
    \includegraphics[width=1\linewidth]{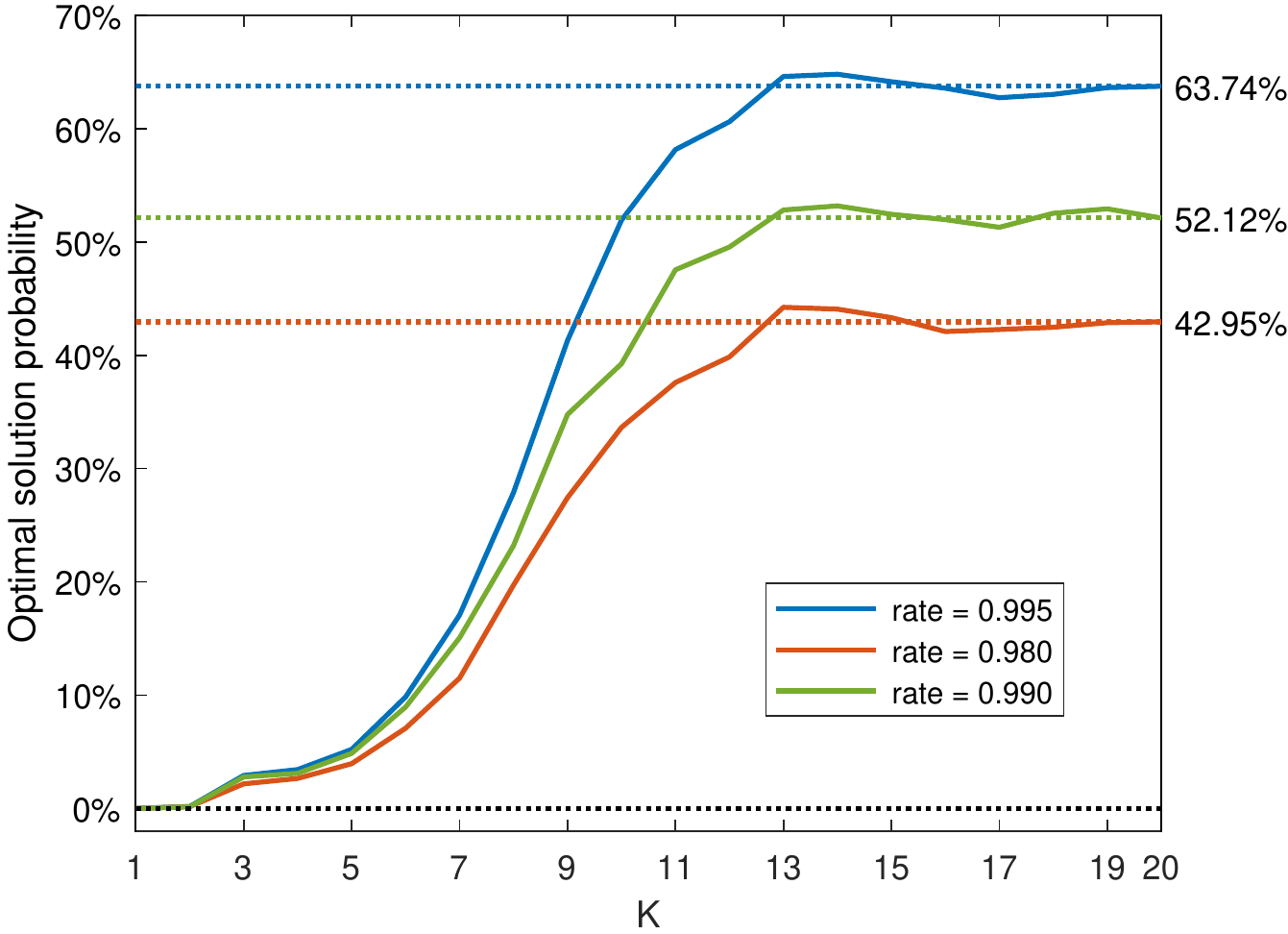}
    \caption{\label{osp} The variation of the optimal solution probability with $K$ for three different simulated annealing parameter settings. They have different cooling schedules, distinguished mainly by the cooling $rate$.}
\end{figure}

The capability of finding the optimal solution is one of the most important metrics when it comes to solve combinatorial optimization problems.
Similarly, for our G-SIM, the optimal probability of the solution at different values of $K$ is a more direct and effective evaluation metric, provided that the optimization algorithm remains the same.
Fig.~\ref{osp} demonstrates the relationship between the probability of obtaining the optimal solution with the value of $K$ for three different cooling schedules of the SA algorithm for the Max-cut problem of Fig.~\ref{firgraph}(a) with random weights. 
The cooling schedules are mainly distinguished by different cooling $rate$s \cite{van1987simulated}, which means that the number of spins that are flipped and the probability of flipping at each iteration gradually decreases as the temperature decreases.
For comparison purpose, the optimal solution probabilities are labeled as dashed lines parallel to the x-axis for $K = 20$, i.e., when HRV represents the Hamiltonian theoretically without error.
Apparently, the probability remains approximately optimal when the value of $K$ is greater than 12, i.e., when approximately 65\% or more of the intensity vectors are modulated.
There are even cases where the probability is slightly higher when part of the intensity vector is assigned ($K<20$). 
In addition, when $K$ is between 5 and 12, the result decreases smoothly with decreasing of $K$, while it keeps at  a relatively low level of 5\% when $K$ is between 3 and 5.
Until $K$ is equal to or less than 2, the algorithms under all three cooling schedules completely fail to find the optimal solution.

\subsection{Discussion}

Firstly, the optimal solution probability shown in Fig.~\ref{osp} is of most interest to us, since finding the optimal solution often implies the successful solution of the combinatorial optimization problem. 
When the intensity vector used for amplitude modulation is below 60\%, the optimal solution probability starts to decrease until it becomes almost zero at $K = 2$. 
The optimal solution of G-SIM with low probability of $K$ configuration seems to be relatively unreliable or difficult to obtain a solution with high confidence.
However, for some practical combinatorial optimization problems, the optimal Hamiltonian is known in advance. 
Or, as long as the optimal solution can be obtained within a finite number of practices, then it can be quickly determined whether it is optimal by verifying its Hamiltonian.
For such problems, the value of $K$ can be reduced even more, since the probability is not zero as long as $K$ is greater than or equal to 3. 
In this case, only a minimum of about 20\% of the intensity vector needs to be modulated, and G-SIM is effective, reducing the implementation cost even more significantly.

Moreover, the mathematical simulation results shown in Fig.~\ref{osp} are obtained for the Max-cut problem for graphs of fixed structure and scale (but with random weights).
The data in Fig.~\ref{osp} are covered for different structures and scales of the plots, and we have found that the RMSE has a similar pattern for different $N$ and $d$, i.e., it decreases exponentially to 0 with $\frac{K}{N}$ according to Eq.~(\ref{exp}).
Combined with the features of matrix spectral decomposition, we reasonably conjecture that the optimal solution probability has a similar performance in the optimization of the Ising model for different structures and scales.
If $N$ is variable, the probability of finding the optimal solution is related to $\frac{K}{N}$.
Hence, for the Ising model of any structure and scale, we consider the optimal solution probability to remain at the optimal level when $\frac{K}{N}$ is kept above 65\%, and the rest is based on the same pattern shown in Fig.~\ref{osp}.

\subsection{Experimental implementation}

\begin{figure}[tb]
    \centering
    \includegraphics[width=0.93\linewidth]{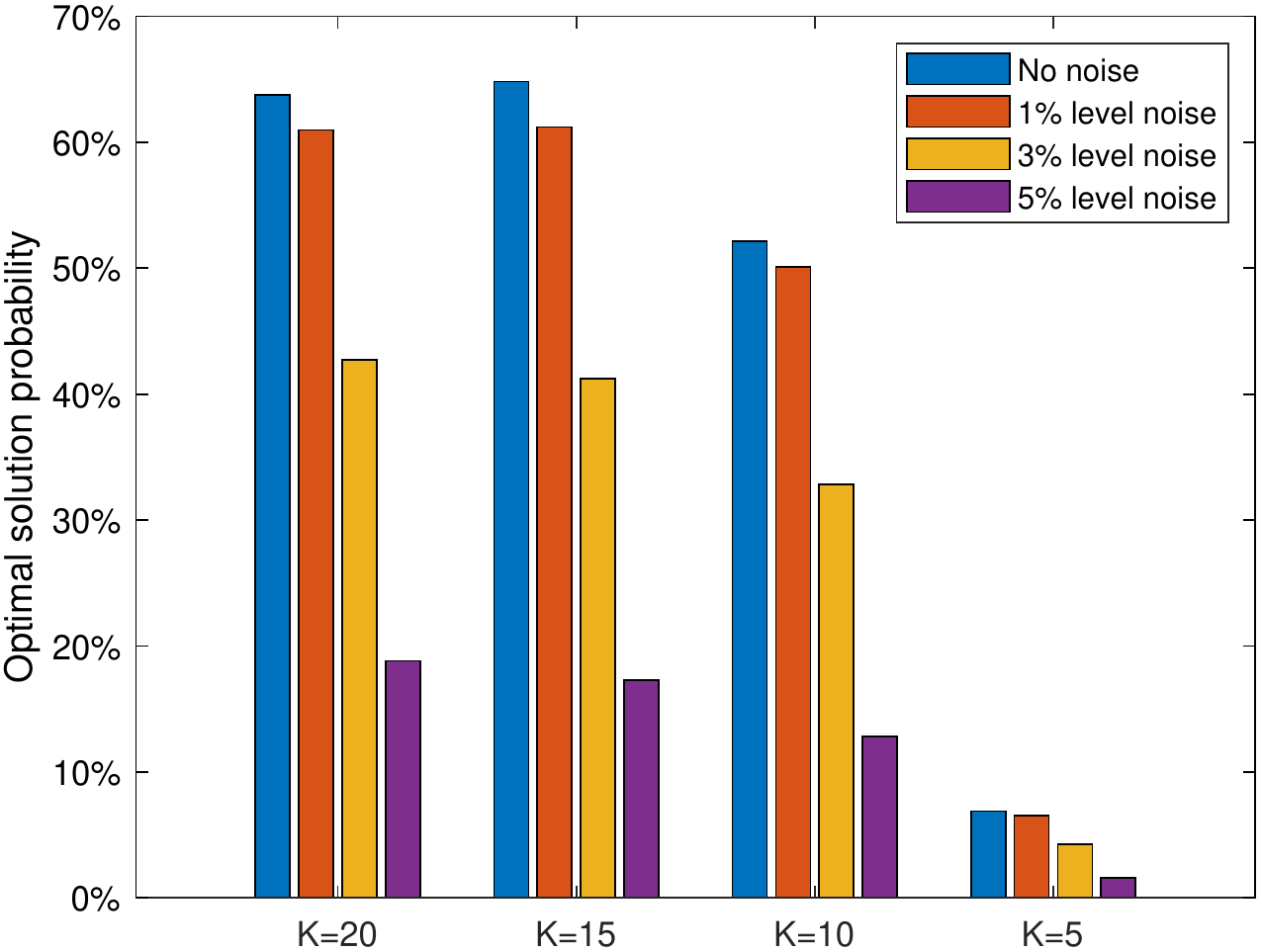}
    \caption{\label{noise} The optimal solution probability obtained by the simulated annealing algorithm with a cooling rate of 0.995 for different noise levels and different K. The percentage value of the noise level implies the ratio of the standard deviation of the added Gaussian noise to the span of HRV values taken.}
\end{figure}

The experiment structure, as illustrated in Fig.~\ref{blockdia}, contains typical optical spatial Ising machine including a low-noise laser, spatial light modulator (SLM), Fourier lens, charge-coupled device (CCD) and I/O that is connected to a processing unit for implementing the spin flipping and iteration control. Thanks to our proposal of the spatial-Euler Ising machine \cite{Ye:23}, 
amplitude modulation and phase modulation can be accomplished by a single SLM.
The Fourier transform is realized by a lens and the detection of the optical field intensity is accomplished by placing the sensor in the CCD on its back focal plane. The spin status is updated through a designed optimization algorithm.

The crucial aspect of the G-SIM experiment is to establish the practical HRV with low error. 
The HRV will inevitably have errors due to hard-to-eliminate precision limitations and external perturbations, as well as operational uncertainties such as beam alignment.
Considering the error introduced by the value of K above and its effect on the optimization results, we believe that acceptable probabilities of optimal solutions will be obtained equally well in the presence of errors introduced by the experimental implementation.
Further, we analyze the effect of the simulated annealing algorithm with the cooling rate of 0.995 in Section III.B after adding different noise levels, as shown in Fig.~\ref{noise}.
Random Gaussian noise is added to the HRV, and we use the ratio of its standard deviation to the HRV span as the judgment of the noise level.
The results show that the optimal solution probability is minimally affected at noise levels around 1\%; and at around 5\%, the optimal solution probability is still maintained at around 1/3 compared to the noiseless condition.


\section{Conclusion}

In this paper, we propose an general spatial photonic Ising machine based on interaction matrix eigendecomposition method.
Arbitrary interaction matrices and external field vectors can be configured using our structure and combined with appropriate optimization algorithms to solve combinatorial optimization problems.
Moreover, we propose amplitude modulation of part of the intensity vectors obtained from the eigendecomposition instead of all of them, which can significantly save the implementation cost and guarantee the reliability of the results. 
In this case, the error of our Ising machine characterizing the Hamiltonian decreases substantially to 0 as the modulated intensity vectors increases.
For the 20-spin Ising model, the matching results are approximately linear when one-fourth of the intensity vectors are modulated, and the error ratio drops to less than 5\% when three-fourths are modulated.
Furthermore, for the Max-cut problem, the optimal solution probability of the optimization algorithm is maintained at an optimal level when the proportion of the utilized intensity vectors is about 65\%, and this proportion can be even reduced to 20\% for some problems.
Finally, we discuss experimental implementation in predictable times and are confident that that experimental errors do not disrupt the functionality of our structure, which is corroborated by the results of the noise analysis.
Our approach enables a considerable enhancement in the ability of spatial photonic Ising machines to solve arbitrary practical problems at minimal cost and has the potential to facilitate the application and development of optical computing platforms.

\section{Acknowledgement}
This work is supported by the National Key Research and Development Program of China under Grant 2019YFB1802903, National Natural Science Foundation of China under Grant 62235011 and 62175146.

\bibliography{mypaper}

\end{document}